\documentclass[final,12pt,a4paper]{article}

\usepackage{mpladefs}
\newcommand{\catchline}[3]{}

\usepackage[top=3cm,bottom=3cm]{geometry}
\usepackage{graphicx}

\usepackage{units}
\usepackage{braket}
\usepackage{color}
\usepackage{amsmath}

\usepackage{feynmp}
\usepackage{ifthen} 
\usepackage{ifpdf}
\usepackage{hyperref}
\usepackage[all]{hypcap} 
\usepackage{subdepth}

\newboolean{articletitles}
\setboolean{articletitles}{true}
\usepackage{mciteplus}

\newcommand{\Bbar}{\ensuremath{{\kern 0.18em\overline{\kern -0.18em B}}}}
\newcommand{\Dbar}{\ensuremath{{\kern 0.18em\overline{\kern -0.18em D}}}}
\newcommand{\Bz}{\ensuremath{B^0}}
\newcommand{\Bzbar}{\ensuremath{\Bbar^0}}
\newcommand{\BzBzbar}{\ensuremath{\Bz\text{-}\Bzbar}}
\newcommand{\Bs}{\ensuremath{B_{\!s}^0}}
\newcommand{\Bd}{\ensuremath{B_{\!d}^0}}
\newcommand{\Bplus}{\ensuremath{B^+}}
\newcommand{\Bminus}{\ensuremath{B^-}}
\newcommand{\Bdbar}{\ensuremath{\Bzbar_d}}
\newcommand{\Bsbar}{\ensuremath{\Bbar_{\!s}^0}}
\newcommand{\BBbar}{\ensuremath{B\text{-}\Bbar}}
\newcommand{\bbbar}{\ensuremath{b\bar{b}}}
\newcommand{\UpsFourS}{\ensuremath{\Upsilon(4\mathrm{S})}}
\newcommand{\ud}{\mathrm{d}}
\newcommand{\DM}{\ensuremath{\Delta m}}
\newcommand{\DMs}{\ensuremath{\DM_s}}
\newcommand{\DMd}{\ensuremath{\DM_d}}
\newcommand{\DG}{\ensuremath{\Delta\Gamma}}
\newcommand{\DGs}{\ensuremath{\DG_s}}

\newcommand{\GL}{\ensuremath{\Gamma_\text{L}}}
\newcommand{\GH}{\ensuremath{\Gamma_\text{H}}}
\newcommand{\GLs}{\ensuremath{\Gamma_{\text{L},s}}}
\newcommand{\GHs}{\ensuremath{\Gamma_{\text{H},s}}}
\newcommand{\Gs}{\ensuremath{\Gamma_s}}
\newcommand{\Gd}{\ensuremath{\Gamma_d}}
\newcommand{\pt}{\ensuremath{p_\perp}}
\newcommand{\muplus}{\ensuremath{\mu^+}}
\newcommand{\muminus}{\ensuremath{\mu^-}}
\newcommand{\mupmum}{\ensuremath{\muplus\muminus}}
\newcommand{\epem}{\ensuremath{e^+e^-}}
\newcommand{\jpsi}{\ensuremath{J/\psi}}

\newcommand{\fzero}{\ensuremath{f^0}}
\newcommand{\VCKM}{\ensuremath{V_\text{\sc ckm}}}
\newcommand{\Vckm}[1]{\ensuremath{V_{#1}^{\phantom{*}}}}
\newcommand{\Vckmcc}[1]{\ensuremath{V_{#1}^{*}}}
\newcommand{\CP}{{\ensuremath{CP}}}
\newcommand{\KS}{\ensuremath{K}_{S}^{0}}

\newcommand{\calO}{\ensuremath{{\cal O}}}

\newcommand{\Dsplus}{\ensuremath{D_s^+}}
\newcommand{\Dsminus}{\ensuremath{D_s^-}}
\newcommand{\Dplus}{\ensuremath{D^+}}
\newcommand{\piplus}{\ensuremath{\pi^+}}
\newcommand{\piminus}{\ensuremath{\pi^-}}
\newcommand{\Kplus}{\ensuremath{K^+}}
\newcommand{\Kminus}{\ensuremath{K^-}}
\newcommand{\BsToDsPi}{\ensuremath{\Bs\to\Dsplus\piminus}}
\newcommand{\BsToDsPiPiPi}{\ensuremath{\Bs\to\Dsplus\piminus\piplus\piminus}}
\newcommand{\BsToDsLX}{\ensuremath{\Bs\to\Dsplus \ell^- \bar{\nu}_\ell X}}
\newcommand{\BsToDsMuNuX}{\ensuremath{\Bs\to\Dsplus \mu^- \bar{\nu}_\mu X}}
\newcommand{\BsToDsMuNu}{\ensuremath{\Bs\to\Dsplus \mu^- \bar{\nu}_\mu}}
\newcommand{\BdToDMuNuX}{\ensuremath{\Bd\to\Dplus \mu^- \bar{\nu}_\mu
    X}}

\newcommand{\BdToJpsiKs}{\ensuremath{\Bd\to\jpsi\KS}}
\newcommand{\BsToJpsiPhi}{\ensuremath{\Bs\to\jpsi\phi}}
\newcommand{\BsToJpsiFzero}{\ensuremath{\Bs\to\jpsi\fzero}}
\newcommand{\BsToKpKm}{\ensuremath{\Bs\to\Kplus\Kminus}}
\newcommand{\phis}{\ensuremath{\phi_s}}

\newcommand{\aFSd}{\ensuremath{a_\text{fs}^d}}
\newcommand{\aFSs}{\ensuremath{a_\text{fs}^s}}

\newcommand{\Dzero}{\mbox{D0}}
\newcommand{\Abar}{\ensuremath{{\kern 0.18em\overline{\kern -0.18em
A}}}}
\newcommand{\phiMG}{\ensuremath{\phi_{12}}}
\newcommand{\aFS}{\ensuremath{a_\text{fs}}}

\begin{document}

\markboth{Wouter Hulsbergen}
{Constraining new physics in \Bs{} mixing}

\catchline{}{}{}{}{}



\date{\normalsize June 2013}
\title{Constraining new physics in \Bs{} meson mixing}

\author{\normalsize Wouter Hulsbergen\\
  \small The Netherlands institute for sub-atomic physics (Nikhef)\\
}

\maketitle


\begin{abstract}
  Neutral mesons exhibit a phenomenon called flavour mixing. As a
  consequence of a second order weak process the flavour eigenstates
  corresponding to the meson and its anti-meson are superpositions of
  two mass eigenstates. A meson produced in a flavour state changes
  into an anti-meson and back again as a function of time. Such
  flavour oscillations are considered sensitive probes of physics
  beyond the Standard Model.  In this brief review I summarize the
  status of experimental constraints on mixing parameters in the \Bs{}
  meson system.

  \bigskip

  \centerline{\emph{To appear in Mod. Phys. Lett. A.}}
\end{abstract}



\section{Introduction}

Quarks are the fundamental fermions that make up baryonic matter in
the universe. In the Standard Model (SM) of elementary particles
quarks come in six flavours, organized in three families,
\[
  \left( 
    \begin{array}{c} 
      \text{up ($u$)}\\
      \text{down ($d$)}\\
    \end{array}
  \right)
  \quad
  \left( 
    \begin{array}{c} 
      \text{charm ($c$)}\\
      \text{strange ($s$)}\\
    \end{array}
  \right)
  \quad
  \left( 
    \begin{array}{c} 
      \text{top ($t$)}\\
      \text{beauty ($b$)}\\
    \end{array}
  \right) \; .
\]
The up-type quarks (top row) have charge $\tfrac{2}{3} e$ while the
down-type quarks (bottom row) have charge $-\tfrac{1}{3} e$. We denote
these quarks with the symbol $q_i$ where the index refers to the
flavour. Quarks have mirror images called anti-quarks, which we denote
with the symbol $\bar{q}_i$. Their physical properties are identical
to those of the quarks, except that they have opposite quantum numbers
for charge and flavour.

In the SM only the charged weak interaction, mediated by the charged
$W$ boson, can change quark flavour. It leads to couplings of the form
$u \to W^+ d$, where the transition is always between an up-type and a
down-type quark. The strength of the coupling is proportional to the
weak coupling constant and to the elements of a complex unitary matrix
that is called the Cabibbo-Kobayashi-Maskawa (CKM) matrix and is
usually represented as
\begin{equation}
  \VCKM \; = \; \left( \begin{array}{ccc}
      V_{ud} & V_{us} & V_{ub} \\
      V_{cd} & V_{cs} & V_{cb} \\
      V_{td} & V_{ts} & V_{tb} \\
      \end{array}
    \right) \; .
\end{equation}
The off-diagonal elements of \VCKM{} are responsible for transitions
between the three quark families. They are small compared to the
on-diagonal elements, which are all close to unity. For three
generations the \VCKM{} matrix can be parametrized by four real
numbers, namely three rotation angles and one phase. The non-zero
value of this phase is the single source of $\CP$ violation (a
difference between matter and anti-matter) in the quark sector of the
SM.\cite{Kobayashi:1973fv} In 2008 Kobayashi and Maskawa were awarded
a Nobel prize for their explanation of \CP{} violation with this
mechanism. In a sense they predicted the existence of the charm,
beauty and top quark well before their discovery.

Mesons are bound states of a quark $q_i$ and an anti-quark
$\bar{q}_j$. (The top quark does not appear in bound states as it
decays too quickly.) The lowest-energy states of mesons with quarks
with different flavour --- those with no net orbital angular momentum
--- can only decay via the weak interaction and are therefore
meta-stable, with lifetimes in the range of $10^{-13}$ to $10^{-7}$
seconds.

If the $q_i \bar{q}_j$ combination is neutral, a phenomenon occurs
that is called \emph{mixing}: the mass eigenstates of such mesons are
quantum-mechanical superpositions of the flavour eigenstate $q_i
\bar{q}_j$ and the \CP{}-conjugate flavour eigenstate $\bar{q}_i q_j$.
As a result a meson created in a $q_i \bar{q}_j$ state may decay as a
$\bar{q}_i q_j$ state with a probability that changes as a function of
time.  Table~\ref{tab:neutralmesons} lists the average decay times and
oscillation period of the four mesons that are subject to mixing.

\begin{table}[h]
  \tbl{Neutral charm and beauty mesons with approximate mass, lifetime and mixing
    period.\protect\cite{pdg2012}
    In the $K^0$ system there are two states with very different
    lifetime. Below only the lifetime of the short-lived state, the
    $K_\text{short}$, is shown.
  }
  {
    \begin{tabular}{ccccc}
      \hline\hline
      name      & quark content & mass/MeV & lifetime/ps & oscillation period/ps \\
      \hline\hline
      $K^0$     & $s\bar{d}$       &   498        &  90  & 1190 \\
      $D^0$     & $c\bar{u}$      &  1865       &  0.41  & 440 \\
      $\Bd$   & $d\bar{b}$     &  5280       & 1.5  & 12.4 \\
      $\Bs$   & $s\bar{b}$     &   5367      & 1.5  & 0.36\\
      \hline\hline
    \end{tabular}
    \label{tab:neutralmesons}
  }
\end{table}

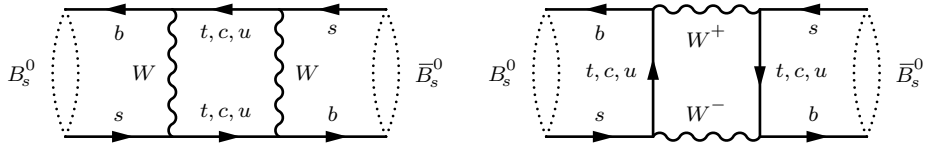
\begin{figure}[htb]
  \centerline{\begin{picture}(350,65)
\scriptsize
\put(20,0) {
  \begin{fmffile}{box1} 
    \begin{fmfgraph*}(120,60)
      \fmfset{dot_len}{1mm}
      \fmfset{arrow_len}{3mm}
      \fmfbottom{i1,d1,o1}
      \fmftop{i2,d2,o2}
      \fmf{fermion,label=$s$,l.side=left}{i1,v1}
      \fmf{fermion,label=$t,,c,,u$,l.side=left}{v1,v2}
      \fmf{fermion,label=$b$,l.side=left}{v2,o1}
      \fmf{fermion,label=$b$,l.side=left}{v3,i2}
      \fmf{fermion,label=$t,,c,,u$,l.side=left}{v4,v3}
      \fmf{fermion,label=$s$,l.side=left}{o2,v4}
      \fmffreeze
      \fmf{boson,label=$W$,l.side=left}{v1,v3}
      \fmf{boson,label=$W$,l.side=right}{v2,v4}
      \fmf{dots,left=0.2}{o1,o2}
      \fmf{dots,left=0.2,label=$\Bsbar$}{o2,o1}
      \fmf{dots,left=0.2}{i2,i1}
      \fmf{dots,left=0.2,label=$\Bs$}{i1,i2}
    \end{fmfgraph*}
  \end{fmffile}
}
\put(200,0){
  \begin{fmffile}{box2} 
    \begin{fmfgraph*}(120,60)
      \fmfset{dot_len}{1mm}
      \fmfset{arrow_len}{3mm}
      \fmfbottom{i1,d1,o1}
      \fmftop{i2,d2,o2}
      \fmf{fermion,label=$s$,l.side=left}{i1,v1}
      \fmf{fermion,label=$b$,l.side=left}{v2,o1}
      \fmf{fermion,label=$b$,l.side=left}{v3,i2}
      \fmf{fermion,label=$s$,l.side=left}{o2,v4}
      \fmf{boson,label=$W^-$,l.side=left}{v1,v2}
      \fmf{boson,label=$W^+$,l.side=left}{v4,v3}
      \fmffreeze
      \fmf{fermion,label=$t,,c,,u$,l.side=left}{v4,v2}
      \fmf{fermion,label=$t,,c,,u$,l.side=left}{v1,v3}
      \fmf{dots,left=0.2}{o1,o2}
      \fmf{dots,left=0.2,label=$\Bsbar$}{o2,o1}
      \fmf{dots,left=0.2}{i2,i1}
      \fmf{dots,left=0.2,label=$\Bs$}{i1,i2}
    \end{fmfgraph*}
  \end{fmffile}}
\end{picture}}
  \caption{Leading order diagrams for neutral meson mixing in the
    SM.}
  \label{fig:boxdiagrams}
\end{figure}

The transition amplitude that governs neutral meson mixing is an
example of a so-called \emph{flavour-changing neutral current} (FCNC).
In the SM neutral meson mixing occurs via a second order weak
amplitude, depicted for \Bs{} mesons in Fig.~\ref{fig:boxdiagrams}. As
such processes are heavily suppressed,\cite{Glashow:1970gm} they are
considered to be very sensitive to contributions from physics beyond
the SM.  Mixing of $\Bs$ mesons is particularly interesting for two
reasons. First, the heavy mass of the $b$ quark allows for relatively
reliable calculations of mixing parameters. Second, it is sensitive to
new contributions in $b\to s$ transitions, which are until now
relatively poorly constrained.

The subject of this review is the status of experimental constraints
on $\Bs$ mixing. We start with a summary of the neutral meson mixing
phenomenology in order to introduce the experimental observables and
SM predictions. Subsequently, experimental techniques and existing
measurements are discussed. We finish with a conclusion and a brief
outlook.

\section{Beauty mixing phenomenology in a nutshell}
\label{sec:theory}

Excellent pedagogical introductions to neutral meson mixing can be
found in textbooks\cite{Branco:1999fs,*Bigi:2000yz}, recent
reviews\cite{Fleischer:2002ys,Lenz:2012mb} and
lecture
notes.\cite{Nierste:2009wg,Buras:2005xt,*Nir:2010jr,*Grossman:2010gw,*Isidori:2013ez}
An up-to-date review of experimental constraints on $B$ meson mixing
can also be found in the PDG.\cite{schneiderpdg2012} The following
discussion applies to neutral mesons of any kind. However, we shall
denote the flavour eigenstate with the symbol $\Bz$ for beauty meson
and use numerical estimates that apply to $\Bs$ and $\Bd$.

\subsection{Time-evolution of the \BzBzbar{} system}

Consider the wave function $\Bz(t)$ for a neutral meson that is the
superposition of flavour eigenstates \Bz{} and \Bzbar{}. The
time-evolution of its projections into flavour eigenstates is given by
a Schr\"odinger equation
\begin{equation}
  i \frac{\ud}{\ud t}
  \left(\begin{array}{c}
      \langle \Bz | B(t) \rangle \\
      \langle \Bzbar | B(t) \rangle \\
    \end{array}
  \right)
  \; = \;
  \left(\begin{array}{cc} H_{11} & H_{12} \\ H_{21} &
      H_{22} \end{array}\right)
  \;
  \left(\begin{array}{c}
      \langle \Bz | B(t) \rangle \\
      \langle \Bzbar | B(t) \rangle \\
    \end{array}
\right) .
\end{equation}
Since the meson decays and we do not consider the wave function of
final states, the Hamiltonian $H$ is not hermitian. However, like any
other complex matrix, it can be decomposed in terms of two hermitian
matrices, which we label by $M$ and $\Gamma$,
\begin{equation}
  H = M - \tfrac{i}{2} \Gamma .
\end{equation}
Since $M$ and $\Gamma$ are hermitian, their diagonal elements are real
and we have $M_{21} = M_{12}^*$ and $\Gamma_{21}=\Gamma_{21}^*$. $CPT$
invariance requires $M_{11}=M_{22}$ and
$\Gamma_{11}=\Gamma_{22}$. Ignoring for the moment the interference
with phases in the final state, the common phase of \Bz{} and \Bzbar{}
is arbitrary such we can choose either the phase of $M_{12}$ or
$\Gamma_{12}$ and only their phase difference matters. Consequently,
the mixing can be parametrized by five real parameters, which are
conventionally chosen to be
\begin{equation}
  M_{11}, \quad \Gamma_{11}, \quad |M_{12}|, \quad |\Gamma_{12}| \quad \text{and} \quad  
  \phiMG =\arg\left( - \frac{M_{12}}{\Gamma_{12}}\right) .
\end{equation}
The mass $M_{11}$ is determined by the quark masses and strong
interaction binding energy. In the $B$ system it is about 5~GeV and
more than ten orders of magnitude larger than the size of the other
elements, which all involve the weak interaction. 

The time-evolution of the meson-anti-meson system is described in
terms of the eigenstates of the Hamiltonian. The two mass eigenstates
can be written as linear combinations of the flavour eigenstates,
\begin{equation} 
  \begin{array}{l}
    | B_L \rangle \; = \; p \:  | \Bz \rangle \; + \; q \: | \Bzbar \rangle\\[2mm]
    | B_H \rangle \; = \; p \: | \Bz \rangle \; - \; q \: | \Bzbar \rangle\\
  \end{array}
\end{equation}
where the subscripts $H$ and $L$ stand for `heavy' and `light', $q$
and $p$ are complex numbers and normalization requires $|p|^2 + |q|^2
= 1$. For $q/p=1$ the mass eigenstates correspond to $\CP$
eigenstates. On the other hand, if $q/p\neq1$, $\CP$ is not conserved
in the time-evolution of the \BzBzbar{} system.

The eigenvalues corresponding to the heavy and light states are
written as
\begin{equation}
  \omega_{L,H} \; \equiv \; m_{L,H} - \tfrac{i}{2} \Gamma_{L,H}
\end{equation}
and are usually recast in terms of the observables
\begin{equation}
  \begin{array}{cc}
    m \equiv \tfrac{1}{2}  \left ( m_H + m_L \right) = M_{11} \qquad & 
    \Gamma \equiv \tfrac{1}{2} \left( \Gamma_H + \Gamma_L \right) =
    \Gamma_{11} \\[2mm]
    \Delta m \equiv  m_H - m_L & 
    \Delta\Gamma \equiv \Gamma_L - \Gamma_H  \: . \\
  \end{array}
\end{equation}
Note that the two eigenstates can have both different mass and
lifetimes. By convention the heavy and light solutions are labeled
such that $\Delta m$ is positive. Using that in the $B$ meson systems
$|\Gamma_{12}| \ll |M_{12}|$ one finds\cite{Nierste:2009wg}
\begin{equation}
  \Delta M            \; \approx \; 2 |M_{12}| \qquad \text{and} \qquad
  \Delta \Gamma \; \approx \; 2  |\Gamma_{12}|
  \cos\phiMG \: .
  \label{equ:G12ToGamma} 
\end{equation}
As we shall see later, $\Delta m$ can be measured by observing an
asymmetry in the decay time distribution of \Bz{} and \Bzbar{}, while
$\Delta \Gamma$ is obtained by combining lifetime measurements of
decays to final states with different $\CP$ content.

The ratio $q/p$ can be written as
\begin{equation}
  \frac{q}{p} \; = \; e^{- i \phi_{M}} 
  \sqrt{ \frac
    { |M_{12}| + \tfrac{i}{2} |\Gamma_{12}| e^{i\phiMG}}
    { |M_{12}| + \tfrac{i}{2} |\Gamma_{12}| e^{-i\phiMG}}
  }
  \label{equ:qOverP}
\end{equation}
where $\phi_{M} \equiv \arg(M_{12})$. For $\phiMG \neq 0, \pi$ the
absolute value of $q/p$ is different from unity, a case we refer to as
\emph{$\CP$ violation in mixing}. Using $|\Gamma_{12}| \ll |M_{12}|$ we
have for the difference of $|q/p|^2$ with one
\begin{equation}
  1 - \left| \frac{q}{p} \right|^2 \; \approx \;
  \left| \frac{\Gamma_{12}}{M_{12}}\right| \sin\phiMG
\end{equation}

Since the time-evolution of the mass eigenstates follows
$\ket{B_{H,L}(t)} = \exp(-i\omega_{H,L}) \ket{B_{H,L}(0)}$, the
time-evolution of a \Bz{} meson produced in a \Bz{} or \Bzbar{}
flavour eigenstate at $t=0$ can now be written as
\begin{equation}
  \begin{split}
    \ket{\Bz(t)} & = \; g_{+}(t) \ket{\Bz} \; + \; \frac{q}{p} \: g_{-}(t)
    \ket{\Bzbar}  \\[2.5mm]
    \ket{\Bzbar(t)} & = \; g_{+}(t) \ket{\Bzbar} \; + \; \frac{p}{q}
    \: g_{-}(t)
    \ket{\Bz}
  \end{split}
\end{equation}
with the functions $g_\pm(t)$ defined as
\begin{equation}
  g_\pm(t) \; = \; \tfrac{1}{2} 
  \left( 
    e^{-i \omega_L } \pm 
    e^{-i \omega_H}
  \right)
\end{equation}
The probability to observe at time $t$ the decay into a state with a
flavour that is the same as (plus sign) or opposite to (minus sign)
the flavour with which is was produced is then proportional to
\begin{equation}
  |g_\pm(t)|^2 \; = \; \frac{e^{-\Gamma t}}{2} \; 
  \Big[ \: \cosh\big(\tfrac{1}{2}\DG\, t \big) \: \pm \: \cos\big( \DM\, t\big) \:
  \Big] ,
  \label{equ:mixingrates}
\end{equation}
while the time-integrated oscillation probability is given by
\begin{equation}
  \chi \; 
  = \; \frac{1}{2} \; \frac{4 \, \DM^2 + \DG^2}{4 \, \DM^2 +
    \Gamma^2} \: .
  \label{equ:chi}
\end{equation}

\subsection{Including decay amplitudes}

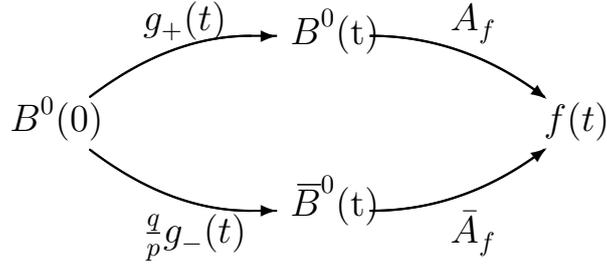
\begin{figure}[htb]
  \centerline{
  \begin{picture}(220,110)
  \large
    \put(0,47){\Bz(0)}
    \put(105,80){\Bz(t)}
    \put(105,15){\Bzbar(t)}
    \put(200,47){$f(t)$}
    \thicklines
    \qbezier(30,60)(60,83)(90,83)
    \put(90,83){\vector(1,0){10}}
    \qbezier(30,40)(60,17)(90,17)
    \put(90,17){\vector(1,0){10}}
    \put(50,85){$g_+(t)$}
    \put(50,5){$\frac{q}{p}g_-(t)$}
    \qbezier(135,83)(170,83)(200,60)
    \qbezier(135,17)(170,17)(200,40)
    \put(200,60){\vector(4,-3){1}}
    \put(200,40){\vector(4,3){1}}
    \put(165,85){$A_f$}
    \put(165,5){$\bar{A}_f$}
  \end{picture}
}
  \caption{Graphical illustration of the interference of two
    amplitudes leading to time-dependent \CP{} violation in decay of a
    neutral meson to a \CP{} eigenstate.}
  \label{fig:mixinginducedcpv}
\end{figure}

The formalism above only describes the time-evolution of the
\BzBzbar{} system and not yet the decay to an observable final state
$f$.  For a given final state we define two transition amplitudes
\begin{equation}
  A_f \; \equiv \; \bra{f} {\cal H} \ket{\Bz}\qquad \text{and} \qquad
  \Abar_f \equiv \; \bra{f} {\cal H} \ket{\Bzbar} \: ,
\end{equation}
where ${\cal H}$ is the weak interaction Hamiltonian responsible for
the decay.  For a meson produced in an initial flavour eigenstate
\Bz{} the decay width to the final state $f$ receives two
contributions, namely one from $A_f$ and another one from $\Abar_f$
where the $\Bz$ first oscillated to a $\Bzbar$, schematically depicted
in Fig.~\ref{fig:mixinginducedcpv}. The partial decay rate can be
written as\cite{Dunietz:2000cr}
\begin{multline}
  \Gamma_{B\to f} (t) \; =  \;
     |A_f|^2 \: \left( 1+ |\lambda_f|^2 \right)  \: \frac{e^{-\Gamma t}}{2} \\
    \Big[ \: \cosh\big(\tfrac{1}{2} \DG\, t \big) 
    + D_f \sinh\big(\tfrac{1}{2}\DG \, t \big) 
    + C_f \cos \big( \DM\, t \big) 
    - S_f \sin \big( \DM\, t\big)  \: \Big]
  \label{equ:decaytimedistribution}
\end{multline}
where we defined
\begin{equation}
  \lambda_f \; = \; \frac{q}{p} \frac{\Abar_f}{A_f}
\end{equation}
and~\footnote{There exist alternative notations for these three
  quantities, in particular $A_f^\text{dir}\equiv C_f$,
  $A_f^\text{mix}\equiv - S_f$ and $A_f^{\DG}\equiv D_f$. However, be
  aware that different conventions are used regarding the signs of
  these quantities.}
\begin{equation}
  C_f \;  \equiv \; \frac{1 - |\lambda_f|^2}{1 + |\lambda_f|^2} \qquad
  S_f \;  \equiv \ \frac{2 \Im(\lambda_f)}{1 + |\lambda_f|^2} \qquad
  D_f \; \equiv \; - \frac{2 \Re(\lambda_f)}{1 + |\lambda_f|^2} \: .
\end{equation} 
The decay rate for $\Bzbar\to f$ is obtained from this expression by
changing the sign of $C_f$ and $S_f$ and multiplying by an overall
factor $|p/q|^2$. Similar expressions can be derived for the decay of
$\Bz$ to the $\CP$ conjugate state $\bar{f}$ by a suitable
redefinition of $\lambda$. It is important to note that, in contrast
to $M_{12}$, $\Gamma_{12}$ and the elements of the CKM matrix,
$\lambda_f$ is a phase convention-independent physical observable.\cite{Carter:1980tk,*Bigi:1981qs}

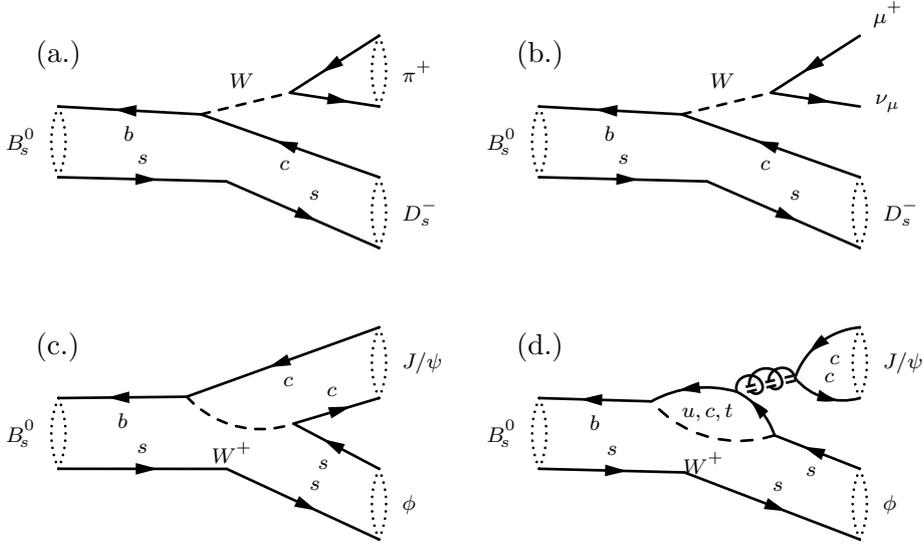
\begin{figure}[htb]
  \centerline{
\begin{picture}(350,210)
\scriptsize
\put(20,110) {
    \begin{fmffile}{dspitree} 
      \begin{fmfgraph*}(120,80)
        \fmfset{dot_len}{1mm}
        \fmfset{arrow_len}{3mm}
        \fmfstraight
        \fmfleft{i1,i2,i3,i4}
        \fmfright{o1,o2,o3,o4}
        \fmf{fermion,label.side=left}{o4,v3}
        \fmf{fermion,label.side=left}{v3,o3}
        \fmf{fermion,tension=2.5,label.side=left,label=$b$}{v2,i3}
        \fmf{fermion,label=$c$,label.side=left,tension=1}{o2,v2}
        \fmf{dashes,tension=2,label=$W$,label.side=left}{v2,v3}
        \fmffreeze
        \fmf{phantom,tension=0.2}{v2,v1,v3}
        \fmf{fermion,tension=0.5,label=$s$,label.side=left}{v1,o1}
        \fmf{fermion,tension=0.5,label.side=left,label=$s$}{i2,v1}
        \fmf{dots,right=0.2}{i2,i3}
        \fmf{dots,left=0.2,label=$\Bs$}{i2,i3}
        \fmf{dots,right=0.2,label=$\Dsminus$}{o1,o2}
        \fmf{dots,left=0.2}{o1,o2}
        \fmf{dots,right=0.2,label=$\piplus$}{o3,o4}
        \fmf{dots,left=0.2}{o3,o4}
      \end{fmfgraph*}
    \end{fmffile}
}
\put(200,110) {
    \begin{fmffile}{dslnutree} 
      \begin{fmfgraph*}(120,80)
        \fmfset{dot_len}{1mm}
        \fmfset{arrow_len}{3mm}
        \fmfstraight
        \fmfleft{i1,i2,i3,i4}
        \fmfright{o1,o2,o3,o4}
        \fmf{fermion,label.side=left}{o4,v3}
        \fmf{fermion,label.side=left}{v3,o3}
        \fmf{fermion,tension=2.5,label.side=left,label=$b$}{v2,i3}
        \fmf{fermion,label=$c$,label.side=left,tension=1}{o2,v2}
        \fmf{dashes,tension=2,label=$W$,label.side=left}{v2,v3}
        \fmffreeze
        \fmf{phantom,tension=0.2}{v2,v1,v3}
        \fmf{fermion,tension=0.5,label=$s$,label.side=left}{v1,o1}
        \fmf{fermion,tension=0.5,label.side=left,label=$s$}{i2,v1}
        \fmf{dots,right=0.2}{i2,i3}
        \fmf{dots,left=0.2,label=$\Bs$}{i2,i3}
        \fmf{dots,right=0.2,label=$\Dsminus$}{o1,o2}
        \fmf{dots,left=0.2}{o1,o2}
        \fmflabel{$\mu^+$}{o4}
        \fmflabel{$\nu_\mu$}{o3}
      \end{fmfgraph*}
    \end{fmffile}
}
\put(20,0) {
  \begin{fmffile}{jpsiphitree} 
    \begin{fmfgraph*}(120,80)
      \fmfset{dot_len}{1mm}
      \fmfset{arrow_len}{3mm}
      \fmfstraight
      \fmfleft{i1,i2,i3,i4}
      \fmfright{o1,o2,o3,o4}
      \fmf{fermion,tension=3.5,label.side=left,label=$b$}{v2,i3}
      \fmf{fermion,label=$c$,label.side=left}{o4,v2}
      \fmf{fermion,label=$c$,label.side=left}{v3,o3}
      \fmf{fermion,label=$s$,label.side=left,tension=2}{o2,v3}
      \fmf{dashes,tension=2.4,label=$W^+$,label.side=right,right=0.3}{v2,v3}
      \fmffreeze
      \fmf{phantom,tension=0.3}{v2,v1,v3}
      \fmf{fermion,tension=0.5,label=$s$,label.side=left}{v1,o1}
      \fmf{fermion,tension=0.5,label.side=left,label=$s$}{i2,v1}
      \fmf{dots,right=0.2}{i2,i3}
      \fmf{dots,left=0.2,label=$\Bs$}{i2,i3}
      \fmf{dots,right=0.2,label=$\phi$}{o1,o2}
      \fmf{dots,left=0.2}{o1,o2}
      \fmf{dots,right=0.2,label=$\jpsi$}{o3,o4}
      \fmf{dots,left=0.2}{o3,o4}
    \end{fmfgraph*}
  \end{fmffile}
}
\put(200,0){
  \begin{fmffile}{jpsiphipenguin} 
    \begin{fmfgraph*}(120,80)
      \fmfset{dot_len}{1mm}
      \fmfset{arrow_len}{3mm}
      \fmfstraight
      \fmfleft{i1,i2,i3,i4}
      \fmfright{o1,o2,o3,o4}
      \fmf{fermion,tension=1.8,label.side=left,label=$b$}{v5,i3}
      \fmf{fermion,tension=1.5,right=0.2,label.side=left,label=$\;\;\;\;u,,c,,t$}{v2,v5}
      \fmf{gluon,tension=2}{v4,v2}
      \fmf{dbl_dashes,tension=0}{v4,v2}
      \fmf{fermion,tension=0.3,right=0.2,label.side=left }{v3,v2}
      \fmf{dashes,tension=0.6,left=0.3,label=$W^+$,label.side=left}{v3,v5}
      \fmf{fermion,label=$c$,tension=0.9,right=0.3,label.side=left}{o4,v4}
      \fmf{fermion,label=$c$,tension=0.9,right=0.3,label.side=left}{v4,o3}
      \fmf{fermion,label=$s$,label.side=left}{o2,v3}
      \fmffreeze
      \fmf{fermion,tension=0.7,label=$s$,label.side=left}{v1,o1}
      \fmf{fermion,tension=1,label.side=left,label=$s$}{i2,v1}
      \fmf{phantom,tension=0.4}{v3,v1,v5}
      \fmf{dots,right=0.2}{i2,i3}
      \fmf{dots,left=0.2,label=$\Bs$}{i2,i3}
      \fmf{dots,right=0.2,label=$\phi$}{o1,o2}
      \fmf{dots,left=0.2}{o1,o2}
      \fmf{dots,right=0.2,label=$\jpsi$}{o3,o4}
      \fmf{dots,left=0.2}{o3,o4}
    \end{fmfgraph*}
  \end{fmffile}}
\put(20,180){\small (a.)}
\put(200,180){\small (b.)}
\put(20,70){\small (c.)}
\put(200,70){\small (d.)}
\end{picture}
}
  \caption{Leading order Feynman diagrams for a few \Bs{} decay modes
    relevant for this review.
  }
  \label{fig:decaydiagrams}
\end{figure}

We now consider two cases relevant for this review. Flavour-specific
final states are final states for which $|\Abar_f| \ll |A_f|$ such
that $\lambda \simeq 0$. Important examples are the tree-level transitions
$\BsToDsPi$ and $\BsToDsMuNu$ shown in Fig.~\ref{fig:decaydiagrams}.
Under the assumption of no \CP{}-violation in mixing ($|q/p|=1$) and
no \CP-violation in the decay ($|A_f| = |\Abar_{\bar{f}}|$), we can
derive the following expression for the so-called time-dependent
oscillation (or `mixing') probability,
\begin{equation}
  A_\text{mix}(t) \; \equiv \; \frac{\left(\Gamma_{\Bz \to f}+\Gamma_{\Bzbar \to
      \bar{f}}\right) - \left(\Gamma_{\Bz \to \bar{f}}+\Gamma_{\Bzbar \to
      f}\right) }
{\left(\Gamma_{\Bz \to f}+\Gamma_{\Bzbar \to
      \bar{f}}\right) + \left(\Gamma_{\Bz \to \bar{f}}+\Gamma_{\Bzbar \to
      f}\right) }
\; = \; \frac{ \cos \DM\, t }{ \cosh \tfrac{1}{2} \DG\, t} \;  .
\end{equation}
The two terms in the definition of the asymmetry are usually called
the `unmixed' and `mixed' contribution, respectively.  In
section~\ref{subsec:dms} we discuss measurements of the oscillation
probability with flavour-specific final states.

Alternatively, dropping the requirement on $q/p$ we can form the
following $\CP$ asymmetry
\begin{equation}
  \aFS \; \equiv \; \frac{ \Gamma_{\Bz \to \bar{f}} - \Gamma_{\Bzbar \to f} }{ \Gamma_{\Bz \to \bar{f}} + \Gamma_{\Bzbar \to f} }
  \; = \; \frac{ 1 - | q/p |^4 }{  1 + | q/p |^4  } 
  \label{equ:afsdef}
\end{equation}
which is notably time-independent. This asymmetry is called the
flavour-specific or semi-leptonic asymmetry. (HFAG denotes this
quantity with ${\cal A}_\text{sl}$.) Note the particle and
anti-particle labels in this definition: both are `mixed'
contributions. In the $B$ meson system $|q/p|$ is close to one. To
translate the measurement of $a_\text{fs}$ into constraints on
$\phiMG$ one often uses the relation
\begin{equation}
  \aFS \approx 1 - |q/p|^2 \approx \frac{\DG}{\DM} \tan \phiMG
\end{equation}
One way to measure $\aFS$ is by counting the number of positive
like-sign and negative like-sign muon pairs in events in which both
$b$ quarks decay semi-leptonically,
\begin{equation}
  \aFS \; = \; \frac{N(\mu^+ \mu^+) - N(\mu^- \mu^-)}{N(\mu^+ \mu^+) + N(\mu^- \mu^-)}
\end{equation}
As we shall see in section~\ref{subsec:asl} the measurement of
flavour-specific asymmetry with semi-leptonic and $B\to D \pi$ decays
provides the best constraints on $|q/p|$ in the \Bs{} system.

Next we consider decays to final states $f$ that are (mixtures of)
\CP{} eigenstates, most notably \BdToJpsiKs{} and \BsToJpsiPhi{}.
Figure~\ref{fig:decaydiagrams}c and d show tree- and penguin-level
contributions to the $\Bs\to\jpsi\phi$ amplitude. If a final state $f$
is accessible to both the \Bz{} and the \Bzbar{} meson, then
interference between decay via mixing and decay without mixing gives
rise to $\CP$-violation. More specifically, if all contributing decay
amplitudes carry the same weak phase, then the decay amplitude ratio
can be written as $A_f / \Abar_f= \eta_f e^{2i\phi_\text{D}}$, where
$\eta_f = \pm 1$ is the \CP{}-eigenvalue of the final state and
$\phi_\text{D} = \arg(A_f)$. If one further assumes that \CP{}
violation in mixing is small ($|q/p|\approx 1$), then
\begin{equation}
  \lambda = \eta_f e^{-i\phi_M} e^{2 i \phi_\text{D}} \: .
\end{equation}
where the phase $\phi_M \equiv \arg(M_{12})$ enters via
Eq.~\ref{equ:qOverP}.  The time-dependent \CP-asymmetry can then be
written as
\begin{equation}
  A_\CP(t) \; \equiv \; \frac{ \Gamma_{\Bz\to f} - \Gamma_{\Bzbar \to f}}{
    \Gamma_{\Bz\to f} + \Gamma_{\Bzbar \to f}} \; = \; 
\frac{ \eta_f \sin \phi_f \sin \DM t}{ \cosh \tfrac{1}{2}\DG t - \eta_f
  \cos\phi_f \sinh\tfrac{1}{2} \DG t} .
\end{equation}
where we introduced the commonly used \CP{} violating phase
\begin{equation}
  \phi_f \equiv -\arg(\lambda_f) = \phi_M - 2 \phi_D \: .
  \label{equ:phif}
\end{equation}
By measuring the amplitude of the sinusoid in the time-dependent
asymmetry we constrain the phase $\phi_f$. Since this phase is related
to the phase of $M_{12}$, the \CP{} asymmetry is a direct probe of new
contributions to $M_{12}$.

For decays to \CP-eigenstates with a single contributing amplitude the
phase $\phi_f$ can be directly expressed in terms of elements of
\VCKM{}. In particular, we have for the so-called `golden modes', that
occur through a tree-level $b\to c\bar{c} s$
transition, \begin{equation}
  \begin{array}{lll}
    \Bd \to \jpsi \KS  & : \quad &  \phi^{c\bar c s}_d = 2\beta \\
    \Bs \to \jpsi \phi & : & \phi^{c\bar c s}_s = -2\beta_s 
  \end{array}
\end{equation}
where the CKM phases $\beta$ and $\beta_s$ are defined by\cite{Dib:1989uz}
\begin{equation}
  \beta \; \equiv \;  \arg\left( - \frac{\Vckm{cd}\Vckmcc{cb}}{\Vckm{td}\Vckmcc{tb}} \right) 
  \quad \text{and}\quad
  \beta_s \; \equiv \;  \arg\left( - \frac{\Vckm{ts}\Vckmcc{tb}}{\Vckm{cs}\Vckmcc{cb}}  \right) \: .
\end{equation}
We discuss the measurement of $\phi^{c\bar c s}_s$ with \BsToJpsiPhi{}
and \BsToJpsiFzero{} decays in section~\ref{subsec:phis}.

Finally, we consider lifetimes. The `untagged' decay time distribution
for a final state $f$ can be obtained from
Eq.~\ref{equ:decaytimedistribution} by setting $C=S=0$. The
average decay time (sometimes called the `effective lifetime') is given
by\cite{Fleischer:2011cw}
\begin{equation}
  \tau_f \; = \; 
  \frac{ (1 - D_f) / \GL^2 + (1 + D_f) / \GH^2}
  { (1 - D_f) / \GL + (1 + D_f) / \GH}
  \; = \; \frac{1}{\Gamma} \frac{1 + 2 D_f y + y^2}{(1-y^2)(1 +
    D_f y)} 
  \label{equ:lifetime}
\end{equation}
where $D_f$ was defined above and $y = \DG / 2\Gamma$. For
flavour-specific modes $D_f=0$ while for decays to \CP{}-eigenstates
with a single contribution amplitude it is $D_f = - \eta_f
\cos\phi_f$. In section~\ref{subsec:lifetimes} we shall discuss
constraints on \Gs{} and \DGs{} from various final states.

\subsection{Standard Model predictions}

In the SM the computation of \Bz{} mixing parameters is performed by
evaluating the amplitudes corresponding to the box diagrams shown in
Fig.~\ref{fig:boxdiagrams}. Since quarks are not free particles, these
amplitudes need to be corrected for hadronization effects. The
calculations have been the cumulative effort of many people, over a
period of over twenty years. (See Refs.~\refcite{Beneke:1996gn,
  *Beneke:1998sy, *Beneke:2003az, Ciuchini:2003ww, Lenz:2006hd,Badin:2007bv} and
references therein.)

The latest complete computation of the mixing observables can be found
in Ref.~\refcite{Lenz:2006hd}, with an update of numerical estimates in
Ref.~\refcite{Lenz:2011ti}. The value of $M_{12}$ is obtained from a
calculation of the box diagram in Fig.~\ref{fig:boxdiagrams} with a
virtual top quark in the loop. The result can be expressed
as\cite{Lenz:2006hd}
\begin{equation}
  M^q_{12} \; = \; 
  \frac{G_F^2 m_W^2}{12\pi^2}
  \,
  \left( \Vckmcc{tq} \Vckm{tb} \right)^2
  \,
  S_0\left( \frac{m_t^2}{m_W^2} \right)
  \:
  \eta_B 
  \:
  \hat{B}_{B_q} f_{B_q}^2 m_{B_q} 
\end{equation}
The part of this expression to the left of $\eta_B$ follows from a
computation of the box diagram for free quarks in perturbation
theory. It depends on parameters of the SM, such as the Fermi coupling
constant $G_F$, the CKM matrix elements $\Vckm{ij}$ and the top quark
and $W$ boson masses. The function $S_0(x)$ is a known kinematic
function, called the Inami-Lim function,\cite{Inami:1980fz} and
$S_0(m_t^2/m_W^2)\approx 2.3$.  The numerical factor $\eta_B \approx
0.55$ accounts for QCD corrections. The factors to its right account
for the fact that the quarks are confined in hadrons.  While the $B$
meson mass $m_B$ is just taken from measurements, the decay constant
$f_{B}$ and the bag factor $\hat{B}_B$ are computed using Lattice
gauge theory. (For a recent review see
Ref.~\refcite{Tarantino:2012mq}). The uncertainty on the prediction of
$M_{12}$ is dominated by the theoretical uncertainty in $\hat{B}_B
f_B^2$.

The computation of $\Gamma_{12}$ involves the evaluation of the box
diagram with `on-shell' internal quarks, the dominant contribution
coming from the $b\to c\bar{c} s$ transition. Since the latter is a
tree-level transition, $\Gamma_{12}$ is expected to be less sensitive
to new physics than $M_{12}$ is. It can be written
as\cite{Buras:1984pq}
\begin{multline}
  \Gamma^q_{12} = - \frac{G_F^2 m_b^2}{8\pi^2}
  \,
  \left [ \left( \Vckmcc{tq} \Vckm{tb} \right)^2 + 
    \Vckmcc{tq} \Vckm{tb} \Vckmcc{cq} \Vckm{cb} \calO\!\left(
      \frac{m_c^2}{m_b^2} \right) + \right.\\
\left.
    (V_{cq}^* V_{cb})^2\calO\!\left( \frac{m_c^2}{m_b^2} \right)
  \right] 
  \:
  \eta'_B 
  \:
  \hat{B}_{B_q} f_{B_q}^2 m_{B_q} 
  \label{equ:gamma12}
\end{multline}
where the QCD correction factor $\eta'_B$ is of order unity.  Note
that $\Gamma_{12}$ is proportional to $\hat{B}_B f_B^2$ as well, such
that predictions for the ratio $\Gamma_{12} / M_{12}$ have smaller
theoretical uncertainty than those for $M_{12}$ and $\Gamma_{12}$
separately. Furthermore, since the ratio is proportional to $m_b^2 /
m_W^2\approx 0.005$, we expect $|\Gamma_{12}| \ll |M_{12}|$ in the SM.

The only source of a complex phase in the expressions for $M_{12}$ and
$\Gamma_{12}$ are the CKM matrix elements.  The size of the CKM matrix
elements in Eq.~\ref{equ:gamma12} is such that the term proportional
to $\left( V_{tq}^* V_{tb} \right)$ dominates. Consequently, taking
into account the minus sign in front of $\Gamma_{12}$, the phase
difference between $M_{12}$ and $\Gamma_{12}$ is approximately $\pi$
and $\phi_{12}$ is small in the SM. This also implies, by virtue
of~Eq.~\ref{equ:G12ToGamma}, that $\DG$ is positive: the heavier mass
eigenstate has the smaller decay width.



The SM does not predict the size of coupling constants,
quark masses and the elements of the CKM matrix. Consequently,
theoretical predictions of mixing observables always rely on other
measurements to determine the SM parameters.  The latter are usually
obtained from global fits to the experimental data that do not include
the mixing observables
themselves.\cite{Charles:2004jd,*Charles:2011va,Ciuchini:2000de,*Bona:2007vi}


\begin{table}[htb]
  \tbl{Predictions for mixing observables taken
    from Ref.~\protect\refcite{Lenz:2011ti}, except for (*), taken
    from Ref.~\protect\refcite{Lenz:2010gu}. }
  {
    \renewcommand{\tabcolsep}{4mm}
    \renewcommand{\arraystretch}{1.3}
    \begin{tabular}{l|c|c}
      \hline
      {}                              & $\Bd$ & $\Bs$ \\
      \hline
      $\phi_{12}$ [rad.]     & $-0.075 \pm 0.024$ & $0.004 \pm 0.001$ \\
      \DG  [ps$^{-1}$]       & $(2.7 \pm 0.5) \cdot 10^{-3}$ & $0.087 \pm
      0.021$ \\
      \DM [ps $^{-1}$]       &  $0.555 \pm 0.073$ *  & $17.3 \pm
      2.6$ \\
      $\aFS$ & $-(4.1 \pm 0.6) \cdot 10^{-4}$ & $(1.9\pm0.3) \cdot
      10^{-5}$ \\[1mm]
      $\phi^{c\bar{c}s}$ [rad.] & $0.84 \pm 0.05$* & $-0.036 \pm 0.002$* \\ 
      \hline
    \end{tabular}
    \label{tab:predictions}
  }
\end{table}

The SM predictions are summarized in table~\ref{tab:predictions}. Note
that, for some results in the table, experimental measurements are
used to reduce the uncertainties. For example, the ratio $\DG_d/\DM_d$
has smaller uncertainties than $\DG_d$ by itself. Consequently, for
the prediction of $\DG_d$, the measured value of $\DM_d$ was used.

Likewise, the computation of \DMs{} requires an estimate of
$\Vckm{ts}$. Since the computation of the perturbative parts of the
amplitude is identical for \Bs{} and \Bd{}, the ratio $\DMs/\DMd$ can
be written as
\begin{equation}
  \frac{ \DMs} {\DMs} \; = \; \xi^2 \frac{ m_{\Bs} } {m_{\Bd}} 
 \left| \frac{ \Vckm{ts} }{\Vckm{td}} \right|^2 ,
\end{equation}
where the so-called $SU(3)$-breaking ratio is defined as
\begin{equation}
  \xi \; \equiv \; \frac{ f_{\Bs} \sqrt{ \hat{B}_{\Bs} }}{ f_{\Bd} \sqrt{
      \hat{B}_{\Bd} }} \: .
\end{equation}
The computation of $\xi$ with lattice QCD has meanwhile reached a
precision of a few
percent.\cite{Laiho:2009eu,*Bazavov:2012zs,*Carrasco:2012dd} With
these expressions one can either test the prediction of \DMs{} by
using other constraints to estimate $\Vckm{ts}/\Vckm{td}$ (as was done
to obtain the results in the table), or use the measurement of
$\DMs/\DMd$ to obtain a precise determination of
$\Vckm{ts}/\Vckm{td}$.

There exists no SM predictions for the $B$ meson masses as these are
essentially determined by the quark masses. Although the total decay
widths cannot be computed reliably either, their ratio is well
constrained and is almost unity in the
SM\cite{Neubert:1996we,Gabbiani:2004tp,Lenz:2011ti}
\begin{equation}
  0 \; \leq \; \frac{\Gs}{\Gd} - 1 \; \leq \; 4 \cdot 10^{-4}.
  \label{equ:lifetimeratioprediction}
\end{equation}
As we shall see, this prediction agrees well with experimental
data. Since uncertainties in this computation rely on similar
assumptions as those for $\Gamma_{12}$ the agreement with data is
sometimes taken as a sign that the computation of $\Gamma_{12}$ is
reliable.

The interpretation of the observable $\phi_f$ in \BdToJpsiKs{} and
\BsToJpsiPhi{} decays in terms of the tree-level quantity
$\phi^{c\bar{c}s}$ ignores a small contribution from penguin decays
with a different weak phase, such as depicted in
Fig.~\ref{fig:decaydiagrams}d.
These contributions may change $\phi_f$ by up to a few degrees and can
be constrained using measurements from other decay modes that are
related by flavour symmetries.\cite{Fleischer:1999nz,*Ciuchini:2005mg,*Faller:2008zc,*Faller:2008gt,*Chiang:2009ev,*DeBruyn:2010hh,*Ciuchini:2011kd,*Fleischer:2012dy,*Bhattacharya:2012ph}


\subsection{Beyond the Standard Model}

Many viable TeV-scale extensions of the SM predict new
flavour-changing neutral couplings, which in turn affect the mixing
parameter $M_{12}$ and for instance the rare decay
$\Bs\to\muplus\muminus$.  As different models affect these quantities
differently, it is through a combination of flavour measurements that
one hopes to identify the correct theory. (For an overview, see
Ref.~\refcite{Buras:2010wr}.)

The effect of new contributions to $M_{12}$ is usually parametrized by
introducing a complex parameter $\Delta_q$ such
that\cite{Grossman:1997dd,Ball:2006xx,Lenz:2006hd} 
\begin{equation}
  M_{12}^q \; \equiv \; M_{12}^{q,\text{SM}} \: \Delta_q
\end{equation}
If the magnitude of $\Delta_q$ differs from unity, this affects the
observed mixing frequency. If its phase differs from zero, this
affects $\phi_{12}$ and thereby $\DG$, $a_\text{fs}$ and the \CP{}
phases extracted from measurements of time-dependent \CP{}
violation. For recent evaluations on constraints on $\Delta_q$ from
\Bs{} mixing measurements, see for instance
Ref.~\refcite{Lenz:2012az}. A recent overview of implications and
relations to other flavour physics observables can be found in
Refs.~\refcite{Buras:2012dp,Bediaga:2012py}.

\section{Experimental facilities and techniques}

The discovery of the $\Upsilon$ (a $b\bar{b}$ bound state) in decays
to $\mupmum$ by Lederman and collaborators in 1977\cite{Herb:1977ek}
marks the onset of beauty-quark physics. Although the $b$ quark was
first found in a fixed-target experiment, most experimental knowledge
comes from two other types of facilities. The first are $\epem$
colliders with a centre-of-momentum energy tuned to the $\Upsilon(4S)$ resonance, a $b\bar{b}$ state
with a mass just above the $\Bd\Bdbar$ and $\Bplus\Bminus$
threshold. These facilities are usually called $\epem$
$B$-factories. The \bbbar{} cross-section at the $\UpsFourS$ is about
\unit[1]{nb}. The fact that this is about one fourth of the total
hadronic cross-section at this energy allows for very clean
studies of $B$ meson properties.

Early $\epem$ $B$-factories (DORIS, CESR) were operated with symmetric
energy beams. As the $\UpsFourS$ resonance is just above the $\BBbar$
meson threshold, $\BBbar$ pairs are produced practically at rest in
the centre-of-momentum system.  Investigations at $\UpsFourS$
facilities took a big step forward with the advance of the
asymmetric-energy $\epem$ colliders KEKB and PEP-II. These
accelerators operated at the $\UpsFourS$ resonance as well, but with a
positron beam energy roughly half that of the electron beam. The
resulting $B$-meson boost, approximately $\gamma\beta=0.5$ at both
colliders, allows for a measurement of the decay time with sufficient
precision to resolve \Bd{} flavour oscillations. Thanks to the
unprecedented instantaneous luminosity, the asymmetric $B$-factories
have collected samples of approximately $10^9$ $\BBbar$ events. The
next generation $\epem$ $B$-factory, super-KEK, is expected to start
operation in 2015, allowing for an increase in statistics with another
factor 50.

Beauty hadrons have also been extensively studied at high-energy
colliders, both at \epem{} machines (SLD, LEP) and at hadron colliders
(SPS, Tevatron, LHC). At these facilities one profits from a larger
cross-section, albeit at the expense of a poorer signal-to-background
ratio.  For example, with a cross-section of about
\unit[$300$]{$\mu$b}, a total of approximately $10^{12}$ \bbbar{}
pairs have been produced in collisions in the first LHC run
(2010-2012). However, these need to be extracted from an inelastic
background that is a factor 200 larger. As only a fraction of events
can be written to permanent storage, the experiments rely on
signatures such as $\jpsi\to\mupmum$ decays, detached muons or
detached high-\pt{} hadrons to select the events of interest.

Besides the larger cross-section the high-energy facilities have two
advantages: First, all species of $b$ quark hadrons are produced, not
only the \Bd{} and $B_u^\pm$ mesons found at the $\UpsFourS$
resonance.  Approximately 10\% of all $b$ ($\bar{b}$) quarks hadronize
into $\Bsbar$ ($\Bs$) mesons. Although studies of \Bs{} mesons have
been performed by operating at higher $\epem\to\Upsilon$ resonances,
conditions at these higher resonances are not favourable enough to
produce competitive \Bs{} samples. Second, thanks to the much larger
$b$ quark boost, the decay time resolution in high energy colliders
far exceeds that at the $B$-factories. As we shall see below, a good
decay time resolution is essential to resolve \Bs{} flavour
oscillations. Consequently, the study of \Bs{} oscillations is (for
now) only performed at high-energy colliders.

\begin{figure}[htb]
  \centerline{
    \begin{picture}(325,130)
      \put(-20,0){\includegraphics[width=0.6\textwidth]{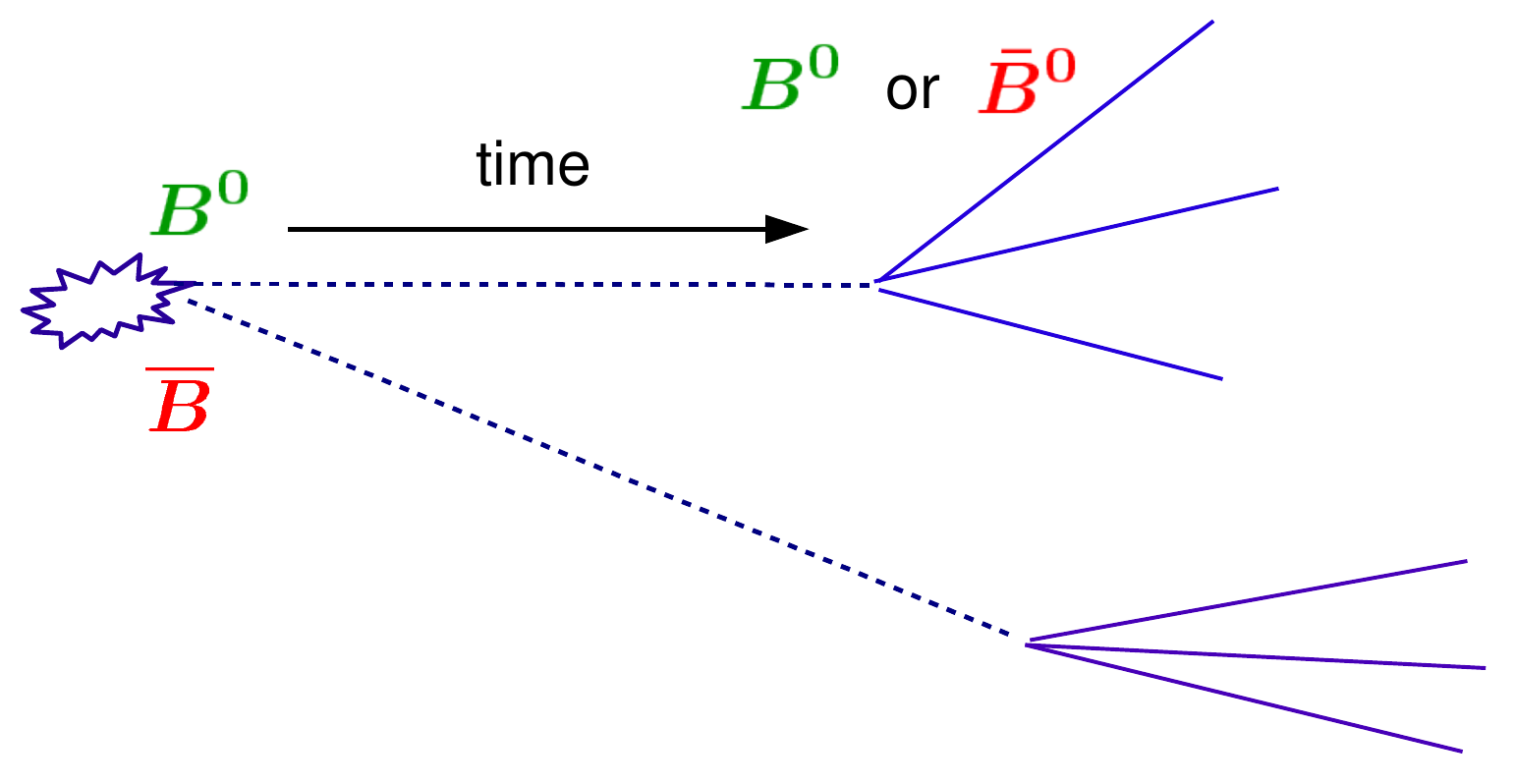}}
      \put(235,90){ \parbox{0.4\textwidth}{
          \small
          \underline{``signal $B$''} :\\
          flavour specific decay\\ or \CP{} eigenstate
        }
      }
      \put(235,15){ \parbox{0.4\textwidth}{
          \small
          \underline{``tagging $B$''} :\\
          determines flavour of\\ signal $B$ at $t=0$
        }
      }
    \end{picture}
  }
  \caption{Schematic representation of the production and decay of a
    $B$ meson in a high-energy collision.}
  \label{fig:measureddecay}
\end{figure}

Figure~\ref{fig:measureddecay} shows schematically the production and
decay of a neutral $B$ meson in a high-energy collider experiment. Due
to the finite decay time of the $B$, its decay vertex is displaced
with respect to the collision point. Silicon vertex trackers have
sufficient position accuracy to measure the decay length $L$, which is
typically a few hundred micron at the \epem{} $B$-factories and up to
centimeters at the hadron colliders.  The decay time in the rest frame
of the particle is then obtained from the observed decay length and
momentum $p$ in the detector frame,
\begin{equation}
  t \; = \; \frac{m L}{p} \: ,
  \label{equ:decaytimemeasurement}
\end{equation}
where $m$ is the rest mass of the particle. At the \epem{}
$B$-factories the production point is not reconstructed and one
measures the difference between the decay time of the ``signal'' $B$
and the ''tagging'' $B$ instead.

\begin{figure}[htb]
  \centerline{\includegraphics[width=0.7\textwidth]{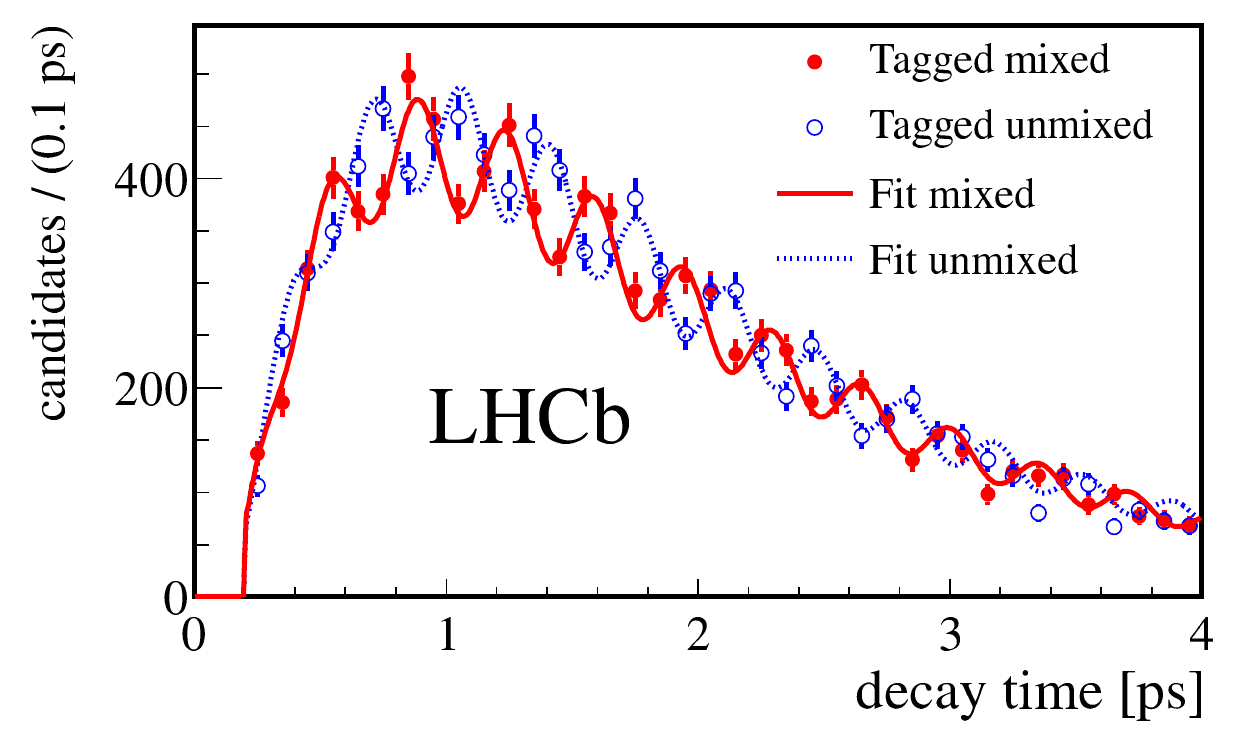}}
  \caption{Observed decay time distribution for samples enhanced in
    `mixed' and `unmixed' \BsToDsPi{} and $\Bsbar\to\Dsminus\piplus$
    candidate decays in 1.0~fb$^{-1}$ of proton-proton collisions at
    the LHCb experiment at CERN.\protect\cite{Aaij:2013mpa}}
  \label{fig:LHCbBsMixing}
\end{figure}

The mixing process described above changes the `flavour' of the meson
as a function of its decay time with a frequency governed by $\DM$. If
one can measure both the flavour at the time of production and the
flavour at the time of decay, the decay time distributions for `mixed'
(equal flavour) and `unmixed' (opposite flavour) events are given by
Eq.~\ref{equ:mixingrates}. The observed distributions for \BsToDsPi{}
decays in an actual experiment -- the LHCb experiment at CERN -- are
shown in Fig.~\ref{fig:LHCbBsMixing}. The mixing frequency \DMs{} is
extracted from the oscillation that modulates the decay time
distribution.

The distribution in Fig.~\ref{fig:LHCbBsMixing}.  differs in several
aspects from the function in Eq.~\ref{equ:mixingrates}. First, at
small decay times, the exponential shape is distorted by
inefficiencies in the event selection. In this particular case, the
drop in efficiency at small $B$ decay times arrises from a requirement
on the minimum distance between the final state tracks and the primary
vertex. This selection was applied in order to remove a large fraction
of the prompt (zero lifetime) background in an early stage of the
event selection. Though less important for the determination of the
mixing frequency, calibration of the decay time acceptance is crucial
to obtain an unbiased estimate of the lifetime parameters $\Gamma$ and
$\DG$.

Second, the observed amplitude of the oscillation is much smaller than
that expected from Eq.~\ref{equ:mixingrates}. This `dilution' of the
oscillation amplitude is caused by the imperfect determination of the
flavour of the initial state and by decay time resolution effects,
both of which we now discuss in more detail.

For flavour-specific final states, such as the \BsToDsPi{} decays used
for Fig.~\ref{fig:LHCbBsMixing}, the flavour at the time of decay
follows from the charge of the final state particles: The charge of
the $D_s$ meson (reconstructed in decays to charged kaons and pions)
uniquely determines whether the $b$ quark was a $b$ or an anti-$b$.

The determination of the flavour at the time of the production of the
\Bz{} is performed with a procedure that is called `flavour
tagging'. Two methods of flavour tagging are used. The first method
relies on the fact that $b$ quarks are produced in $b\bar{b}$
pairs. Consequently, at the time of production there are two $B$
hadrons with opposite $b$ flavour in the event. Assuming that the
flavour of the other $B$ --- called the tag-side or tagging $B$ ---
can be inferred from its decay products, the initial flavour of the
`signal' $B$ follows. The tagging $B$ is usually not fully
reconstructed such that only part of its decay products can be used to
identify its flavour. The charge of high \pt{} leptons and kaons can
be used, as well as the total charge of an inclusively reconstructed
vertex. This procedure is called `opposite-side tagging'. Note that
one inherent limitation to this method is that approximately half the
tagging $B$ hadrons are neutral mesons and hence subject to flavour
oscillations as well, leading to mistakes in the flavour tag.

The second method of flavour tagging exploits that the spectator quark
(the $s$ ($d$) quark for \Bs{} (\Bd{}) mesons) also originates from a
quark-anti-quark pair, leading to a flavour correlation between the
$B$ meson and a light meson close to the $B$ meson in the
fragmentation. The algorithm selects a charged kaon (or pion, for
\Bd{}) that is near in phase space to the $B$ meson. The charge of the
kaon, in case of \Bs{}, or the pion, in case of \Bd{}, reveals the
sign of the $b$ quark. This procedure is usually referred to as
`same-side tagging'.

Mistakes in flavour tagging lead to a dilution of the observed
oscillation asymmetry,
\begin{equation}
  A_\text{mix}^\text{observed}(t) \; = \; D \; A_\text{mix}(t) .
\end{equation}
In practice, the dilution factor $D$ appears in front of all $\sin\DM
t$ and $\cos\DM t$ terms in the differential decay rates, because it is
exactly those terms that change sign between the expressions for an
initial $b$ or initial anti-$b$ state. The dilution factor due to
flavour tagging is equal to
\begin{equation}
  D^\text{tag} = 1 - 2 w
\end{equation}
where $w$ is the fraction of events in which the tag is wrong. Flavour
tagging performance is expressed as the so-called effective tagging
efficiency or tagging power $P = \epsilon_\text{tag} D_\text{tag}^2$
where $\epsilon_\text{tag}$ is the fraction of events for which a
flavour tag could be obtained. Typical values for the tagging power
are \unit[30]{\%} at the \epem{} $B$-factories and a few percent at
hadron colliders.

Besides the imperfect flavour tagging also the effects of finite decay
time resolution lead to a dilution effect. For a Gaussian resolution
$\sigma_t$ the dilution is given by\cite{Moser:1996xf}
\begin{equation}
  D^\text{reso} = \exp\left( - \tfrac{1}{2} \sigma_t^2 \DM^2\right) .
\end{equation}

The decay time resolution at high-energy machines is in general better
than at the $B$-factories due to the larger boost and reduced effects
of multiple scattering. The resolution depends on the final state and
is substantially worse for partially reconstructed decays than for
fully reconstructed decays. For the latter, it ranges from about
\unit[$0.6$]{ps} at the asymmetric-energy \epem{} factories down to
about \unit[$0.05$]{ps} at the LHCb experiment with its forward
geometry. With these numbers the resolution dilution factor for \Bs{}
oscillations is 0.7 at the latter experiment and negligibly small at
the \UpsFourS{} factories, illustrating why the measurement of \DMs{}
is the exclusive domain of experiments at high-energy machines.

For measurements of the $CP$ violating phase $\phi_f$ a proper
calibration of dilution factors is essential. The resolution function
can be obtained from simulations or by taking a process with a known
`zero' decay time distribution, such as prompt $\jpsi$ production. The
flavour tagging performance is calibrated exactly by measuring the
size of the amplitude observed in the decay time distribution of
flavour-specific final states, such as shown in
Fig.~\ref{fig:LHCbBsMixing}.

\section{Status of experimental constraints on \Bs{} mixing}
\label{sec:measurements}

\subsection{Mixing frequency}
\label{subsec:dms}

The first evidence of mixing in neutral $B$ mesons was obtained by the
Argus experiment in 1987: by counting the relative fraction of mixing
and unmixed events the value of the $\Bd{}$ mixing frequency \DMd{}
could be extracted using the expression for the integrated oscillation
probability in Eq.~\ref{equ:chi}.\cite{Albrecht:1987dr} As this was an
integrated rate the decay time of the candidates did not need to be
reconstructed.

For \Bs{} mesons the oscillation period is so small compared to the
lifetime, that the integrated oscillation rate does not provide a
meaningful constraint on the mixing frequency. Rather, the measurement
of \DMs{} is extracted from a fit to the decay time distribution. The
first evidence of mixing in \Bs{} mesons was obtained by the CDF
experiment in 2006, using a combination of fully reconstructed
\BsToDsPi{} and \BsToDsPiPiPi{} and partially reconstructed
semi-leptonic \BsToDsLX{} decays.\cite{Abulencia:2006mq} The current
world average value is dominated by the latest LHCb
result,
\begin{equation}
  \DM_s = 17.768 \pm 0.023 \: \text{(stat)} \pm 0.006 \: \text{(syst)}
  \; \text{[ps$^{-1}$]}
\end{equation}
obtained from the decay time distribution of \BsToDsPi{} events shown
in Fig.~\ref{fig:LHCbBsMixing}.\cite{Aaij:2013mpa} The systematic
uncertainty is determined by the uncertainties in the length scale and
the momentum scale, which enter the measurement of the oscillation
frequency through the decay time measurement in
Eq.~\ref{equ:decaytimemeasurement}.

The current value of $\DMs$ is in good agreement with the SM
predictions presented in section~\ref{sec:theory}.
Assuming the validity of the SM, the value of $\DMs / \DMd$ provides
the best constraint on $\Vckm{ts}/\Vckm{td}$.

\subsection{Measurements of the mixing phase through time-dependent
  \CP{} violation}
\label{subsec:phis}

Measurements of time-dependent \CP{} violation give access to the
mixing phase $\phi_M$. The first such measurements were performed by
Babar and Belle in the golden mode
\BdToJpsiKs{}.\cite{Aubert:2001nu,Abe:2001xe} Their observation of a
large \CP{} violation, in accordance with the prediction in
Table~\ref{tab:predictions}, established the CKM mechanism of the SM
as the dominant source of \CP{} violation in the quark sector.

In the \Bs{} system the best accessible decay channels for this type
of measurement are the decays $\Bs\to\jpsi\phi(1020)$ (with
$\phi(1020)\to\Kplus\Kminus$) and $\Bs\to\jpsi f^0(980)$ (with $f^0
\to \piplus\piminus$). The leading order decay diagrams are shown in
Fig.~\ref{fig:decaydiagrams}.  As explained in
Section~\ref{sec:theory}, the $CP$ violating phase
$\phi^{c\bar{c}s}_s$ is extracted from the amplitude of an oscillation
in the flavour-tagged decay time distribution. Starting from
Eq.~\ref{equ:decaytimedistribution} and assuming $|\lambda_f|=1$ the
latter takes the form
\begin{multline}
  \frac{\ud N_\pm}{\ud t}
  \; = \; N_f \; e^{-\Gamma t} \;
  \Big[
    \cosh\left(\tfrac{1}{2} \DG\, t\right) 
    \; - \; \eta_f \cos\phi_f \sinh\left(\tfrac{1}{2} \DG\, t\right) 
    \\ 
    \; \mp \; \eta_f \sin\phi_f \sin\left( \DM\, t\right)
  \Big]
  \label{equ:tdcpv}
\end{multline}
where the plus (minus) sign on the left-hand-side holds for mesons
produced in a \Bz{} (\Bzbar{}) flavour eigenstate, $\phi_f$ was
defined in Eq.~\ref{equ:phif} and $\eta_f$ is the $CP$ eigenvalue of
the final state.  Final states with $\eta_f=1$ are called $CP$-even,
and those with $\eta_f=-1$ are called $CP$-odd.  For \BsToJpsiPhi{}
and \BsToJpsiFzero{} the observable phase is usually denoted with the
shorthand ``$\phis$'', to distinguish it from the theoretical,
tree-level quantity $\phi^{c\bar{c}s}_s = -2\beta_s$.

The $\jpsi\fzero(980)$ final state is
\CP-odd.\cite{Stone:2008ak,Fleischer:2011au} A recent Dalitz analysis
by LHCb has shown that this holds to a good extend for the entire
$\jpsi\piplus\piminus$ final state.\cite{LHCb:2012ae} The
phenomenology of the $\Bs\to\jpsi\phi(1020)$ decay is more
complicated. Since it concerns a decay to two vector mesons (both the
$\jpsi$ and the $\phi$ have spin one) the final state is a
superposition of states with different angular momentum quantum
numbers, leading to different values of $\eta_f$. In order to extract
\phis{} these contributions need to be statistically disentangled
using the observed decay angles, requiring a so-called `time-dependent
angular analysis' of the data.

Near the $\phi(1020)$ resonance the \BsToJpsiPhi{} final state
receives contributions from four amplitudes, namely three P-wave
amplitudes that belong to the spin-one $\phi\to\Kplus\Kminus$ decays
and a small S-wave component, that is partially from
$f^0(980)\to\Kplus\Kminus$.\cite{Aaij:2013orb} The total decay time
distribution can be written as the sum of 10 equations reminiscent of
Eq.~\ref{equ:tdcpv} above, one for each of the four amplitudes, and
another six for their interference terms.

Although the angular analysis certainly makes the analysis of the data
more complicated, it also has some notable advantages: First, because
of the mixture of odd and even final states, the average observed
decay time is sensitive to both \GH{} and \GL{} (or, equivalently,
$\Gamma$ and \DG{}). Second, as the interference terms contains
flavour-dependent $\sin\DM t$ terms with approximately unit amplitude,
the decay is to a certain extend self-tagging: In principle one can
extract the tagging dilution from the fit to the data without the use
of a tagging control channel. In practise, smaller uncertainties are
obtained if the tagging dilution is constrained to a control sample.
Finally, thanks to these same interference terms, the value of \DMs{}
can be measured. Note that this allows to extract all mixing
parameters from just \BsToJpsiPhi{} decays alone.

The most precise determination of $\phis$ with \BsToJpsiPhi{} and
\BsToJpsiFzero{} events has been reported by LHCb.\cite{Aaij:2013oba}
The two final states yield compatible values, with uncertainties of
about $0.09$ and $0.17$ respectively. The combined result is
\begin{equation}
  \phis \; = \; 0.01 \pm 0.07 \: \text{(stat)} \pm 0.01 \:
  \text{(syst)} \; \text{[rad]} \: ,
\end{equation}
and dominates the current world average. The main contributions to the
systematic uncertainty are from the decay angle acceptance for
\BsToJpsiPhi{} and from the background model in \BsToJpsiFzero{}.

The expression for the partial width Eq.~\ref{equ:tdcpv} remains
invariant under the substitution~$(\DG_q,\phi_q) \longmapsto
(-\DG_q,\pi-\phi_q)$, leaving room for a discrete ambiguity in the
result extracted from the data. The SM predicts $\DG_s >0 $ and
$\phis$ close to zero. In the expressions for \BsToJpsiPhi{} decays,
involving both $P$-wave and $S$-wave amplitudes, the ambiguity
persists, but also involves the (strong) phase differences between the
amplitudes. In particular, the relative phase between the $P$-wave and
$S$-wave amplitude changes sign. Although the relative phase cannot
cleanly be predicted, the variation of the phase difference with the
$\Kplus\Kminus$ invariant mass, which varies rapidly across the
$\phi(1020)$ resonance, is well known,\cite{Wigner:1955zz} allowing
the ambiguity to be resolved.\cite{Aubert:2004cp,Xie:2009fs} Using
this technique measurements by the LHCb collaboration have shown that
only the solution with $\DG_s >0$ and $\phis\approx 0$ is
viable,\cite{Aaij:2012eq} in agreement with the prediction.

\subsection{Lifetimes}
\label{subsec:lifetimes}

Constraints on the average lifetime \Gs{} and the lifetime difference
\DGs{} are obtained by combining information from states with
different \CP{} content, c.f. Eq.~\ref{equ:lifetime}. As indicated
above the analysis of the vector-vector final state \BsToJpsiPhi{}
allows for the extraction of both \Gs{} and \DGs{}. In the absence of
\CP{} violation the lifetime of a \CP{}-odd final decay like
\BsToJpsiFzero{} is $1/\GHs$, while that of the \CP{}-even final state
\BsToKpKm{} is $1/\GLs$. The lifetime measured in flavour specific
decays is equal to $1/\Gs \times (4\Gs^2 + \DGs^2)/(4\Gs^2 - \DGs^2)$.

\begin{figure}[htb]
  \centerline{\includegraphics[width=0.5\textwidth]{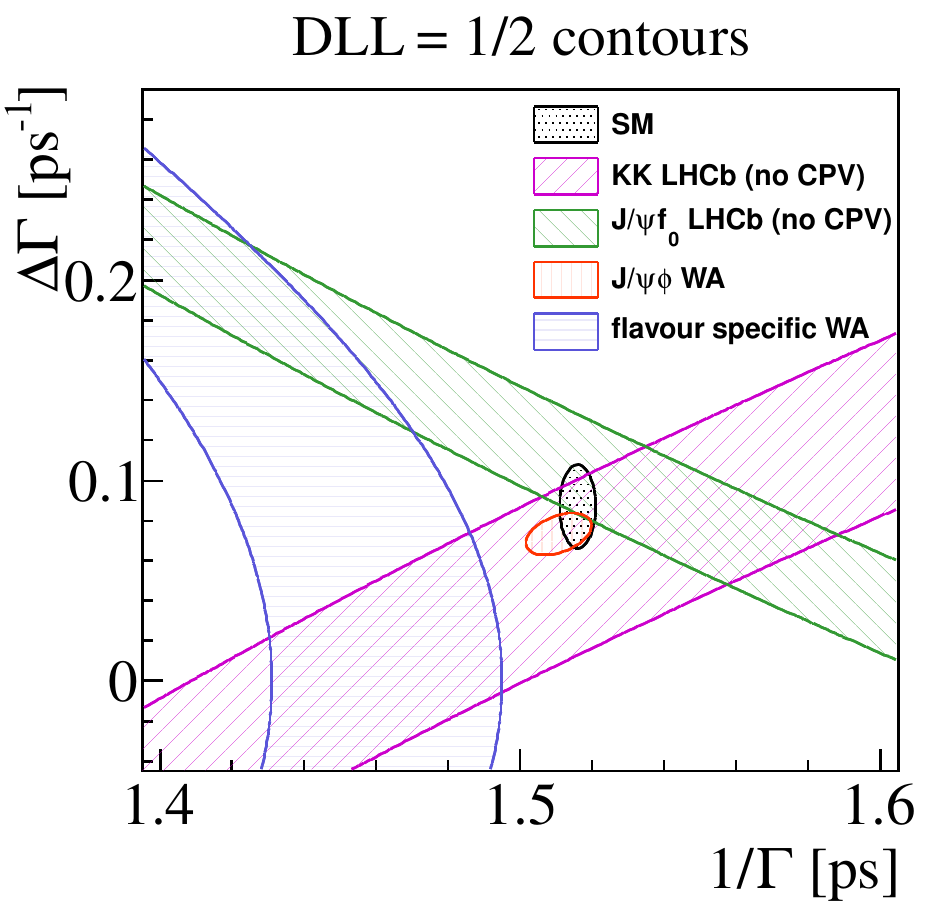}}
  \caption{Constraints on \DG{} and $1/\Gs$ shown as $\Delta\log{\cal
      L}$ contours obtained from the analysis of \BsToJpsiPhi{} decays
    (author's average from D0\protect\cite{Abazov:2011ry},
    CDF\protect\cite{Aaltonen:2012ie},
    ATLAS\protect\cite{Aad:2012kba}, CMS\protect\cite{CMS:2012hya} and
    LHCb\protect\cite{Aaij:2013oba}),
    \BsToJpsiFzero~\protect\cite{Aaij:2013oba},
    \BsToKpKm{}\protect\cite{Aaij:2012ns} and the HFAG average for
    flavour specific decays\protect\cite{Amhis:2012bh}. The
    constraints from \BsToJpsiFzero{} and \BsToKpKm{} are obtained
    under the assumption that there is no \CP{} violation in these
    decays.  The SM prediction (see text) is shown in black.}
  \label{fig:lifetimecontours}
\end{figure}

These different experimental constraints, shown in
Fig.~\ref{fig:lifetimecontours}, are currently in good agreement. The
combined average of the \BsToJpsiPhi{} and \BsToJpsiFzero{} results is
\begin{equation}
  \begin{split}
    \DGs  &= \; 0.081 \pm 0.011 \: \text{[ps$^{-1}$]}\\
    1/\Gs & = \; 1.519 \pm 0.010 \: \text{[ps]}\\
  \end{split}
\end{equation}
The SM prediction, using for \DG{} the value in
Tab.~\ref{tab:predictions} and for $\tau(\Bs)$ the predicted ratio in
Eq.~\ref{equ:lifetimeratioprediction} and the world average value of
$\tau(\Bd)$, is in good agreement with the measurements.  It is
noteworthy that the non-zero value of \DG{} leads to subtleties in the
definition of branching fractions, affecting for instance the
prediction of the $\Bs\to\muplus\muminus$ branching fraction by about
10\%\cite{DeBruyn:2012wj,*DeBruyn:2012wk}.

\subsection{Flavour-specific asymmetry}
\label{subsec:asl}

\begin{figure}[htb]
  \centerline{\includegraphics[width=0.5\textwidth]{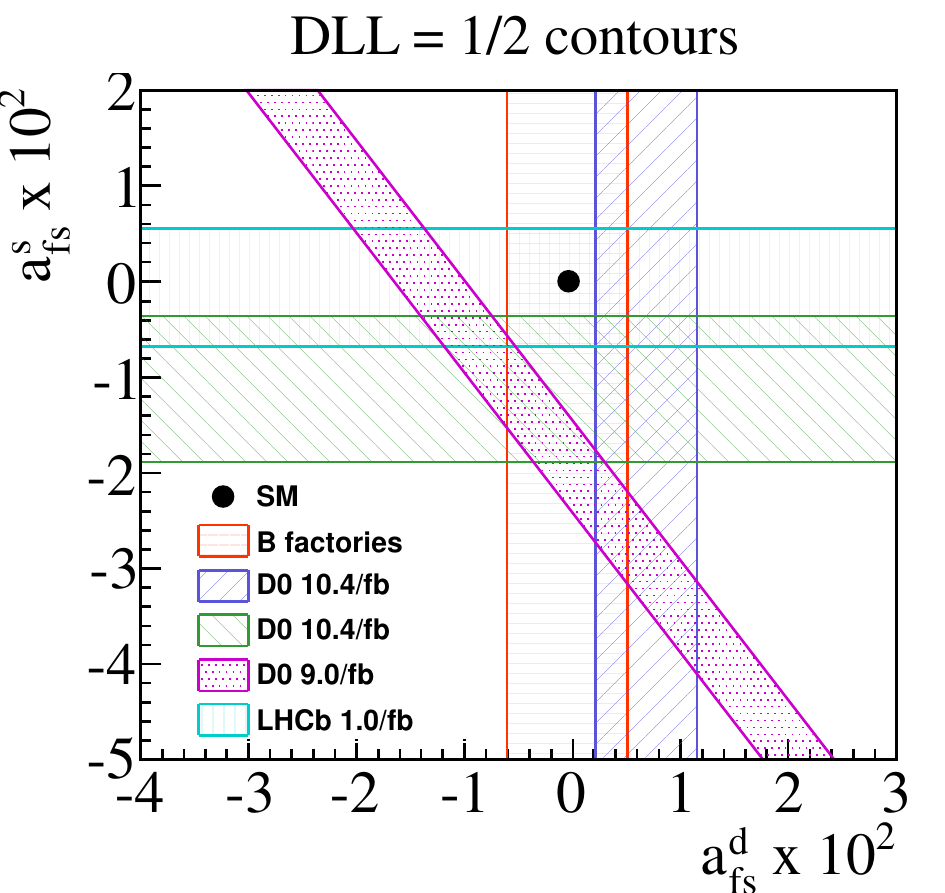}}
  \caption{Constraints on on the flavour-specific asymmetry in the
    \Bd{} and \Bs{} mixing shown as $\Delta \log {\cal L}$ contours
    using data from Babar\protect\cite{Aubert:2006nf},
    Belle\protect\cite{Nakano:2005jb},
    \Dzero{}\protect\cite{Abazov:2012zz,Abazov:2012hha} and
    LHCb\protect\cite{Aaij:2013gta}. The SM prediction is shown
    in black.}
  \label{fig:afs}
\end{figure}

Figure~\ref{fig:afs} summarizes experimental constraints on \CP{}
violation in mixing obtained from measurements of the flavour-specific
asymmetry \aFS{} (defined in Eq.~\ref{equ:afsdef}). A measurement of
\aFS{} requires both an understanding of the relative production rate
of \Bz{} and \Bzbar{} and of the relative reconstruction efficiencies
for the final states $\bar{f}$ and $f$.

At the \epem{} $B$-factories the best constraints in the \Bd{} system
have been obtained using same-sign di-lepton events,
\begin{equation}
  \aFSd \; = \; \frac
  {\Gamma(\UpsFourS \to \ell^+ \ell^+) -\Gamma(\UpsFourS \to \ell^- \ell^-)}
  {\Gamma(\UpsFourS \to \ell^+ \ell^+) +\Gamma(\UpsFourS \to \ell^- \ell^-)} 
  \: .
\end{equation}
The result, $\aFSd= -0.0005 \pm 0.0056$, from a
combination\cite{Amhis:2012bh} of BaBar\cite{Aubert:2006nf} and
Belle\cite{Nakano:2005jb} measurements, is perfectly compatible with
the expectation. A production asymmetry is not a concern at the $B$
factories, but an asymmetry in the efficiency is. This is why these
measurements are systematics dominated, even though they have been
performed with only a fraction of the $B$-factory data set.

As explained above high statistics measurements in the \Bs{} system
can only be performed at high-energy colliders. Two types of probes
have been used. In the same-sign di-lepton analysis \Bd{} and \Bs{}
decays cannot be distinguished, such that at a high-energy collider
one measures a linear combination of the $\aFS(\Bd)$ and $\aFS(\Bs)$,
\begin{equation}
  A_\text{fs}^b \; = \; C_d \, a_\text{fs}^d \; + \; (1 - C_d) \, a_\text{fs}^s
\end{equation}
where the coefficient for \Bd{} is approximately $C_d = 0.59$ at the
Tevatron\cite{Abazov:2011yk}. To reduce the uncertainty from an
eventual tracking efficiency asymmetry \Dzero{} regularly reverses the
magnetic field, a strategy that is also applied by LHCb.  Perhaps the
most tantalizing sign of physics beyond the SM in \Bz{} mixing comes
from the observation of a non-zero value for this effective asymmetry
by the \Dzero{} collaboration\cite{Abazov:2011yk}, shown by the
diagonal band in Fig.~\ref{fig:afs}.

An important concern in the same-sign di-lepton analysis is the
understanding of asymmetries in the non-$B$ backgrounds, for instance
in muons from kaon and pion decays in flight. The purity for the
signal can be substantially be improved with a more exclusive
reconstruction, that also allows for a separation of the \Bd{} and
\Bs{} contribution: the asymmetry in $\BsToDsMuNuX$ production has
been studied by both \Dzero{}\cite{Abazov:2012zz} and
LHCb\cite{Aaij:2013gta}, while \Dzero{} has looked in addition
at the $\BdToDMuNuX$ asymmetry\cite{Abazov:2012hha}.


In these analyses the opposite side $b$ hadron is not tagged, leading
to a slightly different relation to \aFS. Including the effect of an
eventual production asymmetry, defined as
\begin{equation}
  a_\text{prod} \; = \; \frac{N(\Bz(t=0)) - N(\Bzbar(t=0))}{N(\Bz(t=0)) + N(\Bzbar(t=0))}
\end{equation}
the observed asymmetry is related to \aFS{} by
\begin{equation}
  \frac
  { N( {D_q^{-}\mu^+}) -N( {D_q^+ \mu^-} )}
  { N( {D_q^{-}\mu^+}) + N( {D_q^+ \mu^-} )}
  =
  \frac{\aFS}{2} +
  \left( \frac{\aFS}{2} - a_\text{prod}\right)
  \frac{ \int_0^\infty \ud t \: e^{-\Gamma t} \cos(\Delta m \, t)}
  { \int_0^\infty \ud t \: e^{-\Gamma t} \cosh(\tfrac{1}{2}\Delta
    \Gamma \, t)}
   \label{equ:dsmuasym}
\end{equation}
At the Tevatron, a proton-anti-proton collider, the production
asymmetry is zero. At the LHC it is expected to be at the percent
level, but with large uncertainty. As a consequence of the rapid
oscillations, the integral on the right hand side of Eq.~\ref{equ:dsmuasym} is of the order of 1
per mille for \Bs{} mesons, which strongly dilutes any contribution
from the production asymmetry. This is not the case for \Bd{} mesons
and explains why LHCb can measure \aFSs{} but not \aFSd{} with this
method. As shown in Fig.~\ref{fig:afs} the semi-exclusive $D^+_q
\mu^-$ analyses are competitive with the di-lepton analysis. So far
they are in good agreement with the SM.

\section{Concluding remarks and Outlook}

I have presented a brief review of experimental constraints of mixing
phenomena in \Bs{} decays. These phenomena are interesting because
they are cleanly predicted in the SM and sensitive to
physics at higher mass scales.  While mixing in the \Bd{} system has
been extensively studied at the \epem{} $B$-factories, the \Bs{}
system is the exclusive domain of high-energy colliders. The Tevatron
experiments CDF and \Dzero{} have shown the potential of this
research, being the first experiments to observe \Bs{} oscillations
and rule out leading order new physics effects. The LHC experiments,
in particular LHCb, have caught on quickly, providing even stronger
constraints on the oscillation frequency, lifetimes and time-dependent
\CP{} violation.

With only a subset of the data from the first LHC run analysed, and
more luminosity expected in 2015 and beyond, we can expect
significantly tighter constraints in the near future. The increase in
available statistics also allows the study of mixing-induced \CP{}
violation in more decay modes, such as
$\Bs\to\phi\phi$\cite{Aaij:2013qha}, $\Bs\to h^+ h^-$ and $\Bs\to
D^{(*)}\Dbar^{(*)}$.  Clearly, as statistical precision increases,
controlling systematic uncertainties, both experimental and
theoretical, becomes more important.  Experimental uncertainties in
lifetime and \aFSs{} are expected to be dominated by detector effects
rather soon. On the other hand, combinations of \CP{} violation and
branching fraction measurements for different channels, will help to
reduce theoretical uncertainties due to subdominant amplitudes and
non-perturbative effects.

Finally, we can expect another jump in precision near the end of the
decade. The upgrade of the KEKB accelerator and Belle detector are
well under way, with the start of data taking planned for
2015\cite{Abe:2010sj}.  The LHCb collaboration is preparing an upgrade
that allows for a 10-fold increase in integrated luminosity, to be
collected in a five year period starting approximately in
2019\cite{Bediaga:1443882}. These flavour physics facilities are both
competitive with and complementary to the direct searches for new
forces and particles in high energy collisions.


\section*{Acknowledgments}

The author is grateful to Prof. Dr. Gerhard Raven and Prof. Dr. Robert
Fleischer for proofreading the manuscript and providing many useful
suggestions. The author receives funding from the Netherlands
organisation for scientific research (NWO/FOM) through the VIDI scheme.

\bibliographystyle{LHCb}
\bibliography{main}
\end{document}